# A Tutorial for Evaluating Cure Model Appropriateness

Geethanjalee Mudunkotuwa[1*], Durbadal Ghosh[2*], Brandon Triplett[3], Subodh Selukar[2]

[1]Department of Biostatistics & Data Science, University of Kansas Medical Center, Kansas City, Kansas

[2]Department of Biostatistics, St. Jude Children's Research Hospital, Memphis, Tennessee

[3]Department of Bone Marrow Transplantation & Cellular Therapy, St. Jude Children's Research Hospital, Memphis, Tennessee

* denotes equal contribution

Correspondence: Durbadal Ghosh, Department of Biostatistics, St. Jude Children's Research Hospital, 262 Danny Thomas Place, Memphis, TN 38105, USA. Email: durbadal.ghosh@stjude.org

## ABSTRACT

In survival analysis, traditional models assume all individuals will eventually experience the event of interest. However, advances in therapeutics have led to multiple clinical contexts with potentially curative therapies, and in these contexts, certain individuals may never experience the event. Statisticians have developed cure models as a methodology to address this challenge. However, despite significant statistical advances in cure models, we have seen more limited uptake in biomedical applications, and we hypothesize that this reflects limited guidance on the appropriate application of cure models. Cure models require specific identifiability conditions for valid parameter estimation, and previous reports have demonstrated significant issues with the inappropriate application of cure models. Existing tutorials for cure models focus on model implementation and either assume or provide only limited guidance on whether cure modeling is appropriate for the given dataset. This tutorial addresses this gap by describing a systematic procedure that integrates clinical judgment, visual inspection of Kaplan-Meier curves, and quantitative evaluation. We provide a worked example using data from a randomized clinical trial in acute myeloid leukemia, and we also summarize findings from a series of other datasets of hematopoietic cell transplantation to suggest broad practical guidance for choosing to apply cure models. By systematically evaluating cure model appropriateness before fitting these models, researchers can achieve more reliable survival analysis and improved clinical decision-making.

## 1 | INTRODUCTION

### 1.1 | Background

Advances in oncology therapies have led to increased long-term survival in certain cancer populations. For example, recent developments in immune checkpoint inhibitors offer potentially improved progression-free survival for patients in early disease stages[1]. And, while multiple myeloma is generally considered incurable, certain subtypes and early-stage cases can achieve long-term remission with chemotherapy, particularly when combined with stem cell transplantation and novel therapies[2].

Cure models originated in the 1940s[3] and have since been applied across diverse disciplines, from oncology to health economics, credit scoring, and engineering reliability[4-6]. While traditional survival models assume that all individuals are susceptible to an event of interest and will eventually experience it, cure models were developed to analyze contexts in which a non-zero fraction of individuals are cured or will never experience the event of interest. In appropriate contexts, cure models can outperform traditional approaches by providing better prediction of event rates when planning clinical trials[7], more efficient test statistics[8], and improved estimation in long-term extrapolations[9,10].

However, cure models make important assumptions beyond those of traditional survival models. Two key assumptions are that (1) there exists a non-zero proportion of the population that will not experience the event and (2) there exists suitable follow-up in the sample to identify cure model parameters. (The latter condition has received much attention in the literature and is developed with more technical precision in Section 2.1.2.)

Authors have shown that violations to these assumptions can cause significant issues in the results from cure model analysis. Simulation-based studies by Grant et al.[11] and Kearns et al.[12] have shown that cure model estimates derived from interim or limited follow-up can differ substantially from those obtained at later, more mature follow-up, and that model misspecification further degrades extrapolation performance. To evaluate the impact of sufficient follow-up time on parameter estimation with cure models in real-world data, Othus et al.[13] analyzed six clinical trials from the SWOG Cancer Research Network each at two different follow-up times. They observed that initial cure model-based estimates of mean survival could meaningfully differ from later estimates, and that the direction of the difference was not predictable. This research demonstrates that misapplying cure models can produce unreliable results in real-world studies, potentially misleading clinicians and policymakers who evaluate long-term treatment effectiveness.

In recent literature, three tutorials were published on the use of cure modeling, all in the context of health technology assessment (HTA)[14-16]. These tutorials provide comprehensive guidance for applying cure models—including parametric distribution selection by Felizzi et al.[14], excess hazard methods by Sweeting et al.[15], and comparison of mixture versus non-mixture approaches by Latimer and Rutherford[16]. However, they do not address a fundamental prerequisite: whether a given dataset is likely to be appropriately modeled with a cure model (e.g., whether a study's follow-up is sufficiently long to permit valid cure modeling).

In this work, we aim to address this gap with a tutorial that helps users understand both the underlying concepts and the practical steps for identifying whether a cure model is appropriate for a given dataset. First, we introduce the data setup and summarize theoretical conditions for considering cure models. Then we discuss methods of visual evidence for cure fraction and sufficient follow-up followed by quantitative methods of assessing cure model appropriateness. Next, we demonstrate a step-by-step process of assessing cure model appropriateness using a worked example with R code from our new, publicly available cureAssess package. Finally, we describe other practical considerations for applying cure models based on a series of datasets from hematopoietic cell transplantation.

## 2 | METHODS

## 2.1 | Mixture Cure Models

For each subject i = 1, ..., n, let $T_i$ denote the failure time. Subjects are observed under an independent censoring time $U_i$, and the observed data consist of ($Y_i$, $\delta_i$), where $Y_i = \min(T_i, U_i)$ is the observed time and $\delta_i = I(T_i \leq U_i)$ is the event indicator. We assume the failure times $T_i$ are independent and identically distributed (i.i.d.) with cumulative distribution function $F(t) = P(T \leq t)$ and survival function $S(t) = P(T > t)$, $t \geq 0$. In a cure setting, these distribution and survival functions can be improper, with $\lim_{t\to\infty} S(t) > 0$ and $\lim_{t\to\infty} F(t) < 1$, reflecting a non-zero probability that the event of interest is never observed.

In this tutorial, we concentrate on mixture cure models[3,17] for consistency with the literature on cure model appropriateness and their usage in practice relative to non-mixture models. Mixture cure models write the population survival function as a mixture of cured and uncured sub-populations. Let the probability of being in the cured sub-population be $(1 - p)$, and the survival function for the uncured sub-population be the proper survival function $S_u(t)$. The mixture cure model is then:

$$S(t) = (1 - p) + p \cdot S_u(t), \quad t \geq 0 \quad (1)$$

Note that this formulation models cure status as a latent property of the population: an individual's cure status is not directly observed, and the model is best understood as describing the proportions of long-term and short-term survivors in the population rather than as a tool for classifying individual subjects[18].

### 2.1.1 | Practical Considerations when the Outcome of Interest Includes Mortality

In oncology, many time-to-event analyses focus on mortality endpoints, where the literal notion of being "cured" or "immune" to death is unrealistic. For event-free survival endpoints, individuals may avoid some components of a composite outcome (e.g., disease relapse) but not death itself. In these settings, cure models are best understood as capturing long-term survivorship rather than absolute immunity: the cure fraction is interpreted as the proportion of the population whose event hazard becomes negligible after a sufficient duration of follow-up. A cure model can therefore still fit the data well when the long-term sub-population is unlikely to experience the event during the observed follow-up. Additionally, the class of relative survival models[19] is an alternative class that explicitly accounts for background mortality, if needed to address the residual hazard due to background mortality.

### 2.1.2 | Identifiability Conditions for Cure Models

Cure models require identifiability conditions beyond those of conventional survival models. "Identifiability" concerns whether the model parameters can be uniquely determined from observed data, and two types of identifiability[20] are relevant for cure models.

The first type concerns the model structure itself. It asks whether an observed population survival function can be uniquely decomposed into its cure and uncured survival components.

Hanin and Huang[21] defined this form of identifiability with the statement that the model is identifiable within function classes *P* and *S* if

$$(1 - p_1) + p_1 \cdot S_{u,1}(t) = (1 - p_2) + p_2 \cdot S_{u,2}(t) \text{ for all } t \geq 0 \quad (2)$$

implies $p_1 = p_2$ and $S_{u,1} = S_{u,2}$ for $p_1$ and $p_2$ in *P* and $S_{u,1}$ and $S_{u,2}$ in *S*. This type of identifiability depends only on the mathematical form of the model. It determines whether distinct parameter values produce distinct population survival functions, regardless of what data are observed. They then established the conditions under which various parametric and nonparametric specifications of *P* and *S* yield identifiable models.

The authors showed that mixture cure models with common parametric uncured distributions, such as Weibull, log-logistic, gamma, or log-normal, are identifiable in this first sense provided the parametric family of the uncured distribution is itself identifiable. The semiparametric mixture cure model with proportional hazards on the uncured survival (with unspecified baseline hazard) is also identifiable in this sense, typically requiring covariates that distinguish the cure component from the uncured survival component.

The second type of identifiability is directly relevant to parameter estimation from observed data[20]. It requires that a unique set of parameters maximizes the expected log-likelihood given the observed data distribution. Patilea and Van Keilegom[22] established conditions for this under parametric incidence models, while Xu and Peng[23] addressed the nonparametric case. Although some conditions apply to all survival analysis models, these articles introduce additional follow-up assumptions needed to obtain desirable properties for cure models. One representation of this type of assumption characterizes the relationship between the event time distribution and the censoring distribution through the so-called sufficient follow-up condition[24]:

$$\tau_F_0 \leq \tau_G \quad (3)$$

where $\tau_{F_0} = \sup\{t : F_0(t) < 1\}$ denotes the upper support of the event time distribution for susceptibles ($F_0 = 1 - S_u$), and $\tau_G = \sup\{t : G(t) < 1\}$ is the upper support of the censoring distribution $G$. Condition (3) requires that follow-up extends beyond the time at which all susceptible individuals would have experienced the event.

When condition (3) fails, late censored observations could represent either cured individuals or susceptible individuals who would eventually experience the event. These cases are observationally indistinguishable, and the cure fraction cannot be uniquely estimated. Even when the model permits unique decomposition based on the first type of identifiability, estimation requires that condition (3) holds. Methods for evaluating “sufficient follow-up” assess this second form of identifiability, which is essential for valid parameter estimation from observed data.

It is important to distinguish this statistical condition from clinical conventions in which a fixed survival duration (e.g., 5 years post-diagnosis in oncology) is used to designate a patient as "cured." Such conventions reflect domain-specific norms about clinically meaningful long-term survival rather than the identifiability requirement of equation (3). The follow-up needed for valid estimation under a cure model will vary across datasets and must be assessed on a per-study basis using the methods described in Section 2.2.

**2.2 | Assessing Cure Model Appropriateness**

The following sections describe an assessment process for cure model appropriateness, which proceeds through three stages: clinical judgment, visual assessment, and quantitative evidence. Figure 1 summarizes this process. At each stage, evidence may indicate that a cure model is inappropriate. In these situations, statisticians frequently apply traditional non-cure models. However, when cure modeling seems reasonable, but follow-up appears insufficient, authors have recently developed alternative methodology. For example, Escobar-Bach and Van Keilegom[25] and Escobar-Bach et al.[26] developed nonparametric estimators of the cure fraction that remain valid under insufficient follow-up by leveraging extreme-value theory. In this tutorial, we focus on the case where a user is interested in evaluating the appropriateness of a cure model in one group, which has been best studied in the literature, and we discuss multiple groups in the Discussion.

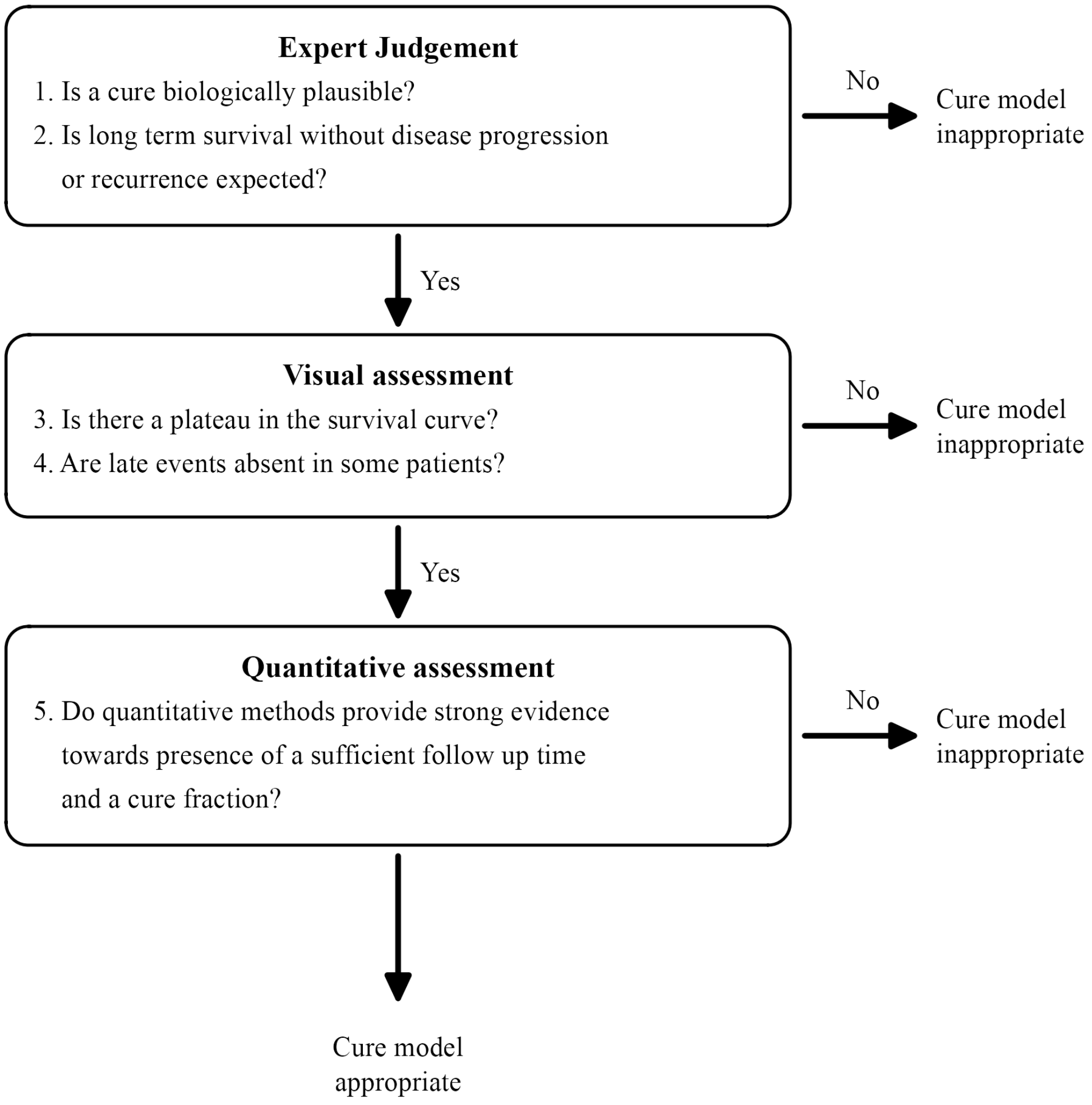

[FIGURE 1: Workflow for evaluating cure model appropriateness]

### 2.2.1 | Step 1: Clinical Judgment

Statisticians should engage clinical experts to evaluate whether cure models could be relevant data models for the specific application. Reviewing prior applications of cure models in comparable datasets can offer guidance, but expert knowledge of long-term patient trajectories remains critical. In particular, expectations regarding the potential for late failures will inform interpretations of follow-up sufficiency. These discussions can also clarify that true cure may be inappropriate (see Section 2.1.1) and review whether modeling long-term survivorship with cure models may be useful.

### 2.2.2 | Step 2: Visual Evidence for Cure Fraction and Sufficient Follow-up

The Kaplan-Meier curve provides a nonparametric estimate of $S(t)$ and serves as the primary tool for visual assessment.

Two features in the Kaplan-Meier curve support evaluating cure model appropriateness. First, if the curve levels off to a horizontal plateau that does not approach zero, this suggests $\lim_{t \to \infty} S(t) > 0$, consistent with a non-zero cure fraction. Second, a decreasing event rate prior to the plateau can suggest a population in which susceptible individuals have largely experienced the event, leaving predominantly cured individuals. Recall that each downward step in a Kaplan-Meier curve represents observed events, with height weighted by the number at risk. Because the number at risk decreases over time, late steps are necessarily larger even when events occur at the same or slower pace. The pattern supporting cure model appropriateness is therefore fewer events per unit time near the end of follow-up, not necessarily smaller steps.

Visual inspections are inherently subjective, which can lead to low confidence in decisions for cure modeling, especially in situations with limited prior knowledge. To provide further confirmation when assessing cure model appropriateness, authors have developed quantitative methods as well.

### 2.2.3 | Step 3: Quantitative Evidence

Since Maller and Zhou introduced the concept of sufficient follow-up in 1992[27], there has been significant focus on developing quantitative methods to evaluate follow-up adequacy. In their textbook, Maller and Zhou[17] described a two-step hypothesis testing procedure to determine whether a cure model should be applied.

Building on the data setup in Section 2.1, let $G$ denote the cumulative distribution function of the censoring time $U_i$, assumed independent of the failure times. In the mixture cure model, the population distribution function can be written as $F(t) = p \cdot F_0(t)$, where $F_0$ is the proper distribution function of the uncured sub-population and $(1 - p)$ remains the cure fraction. Define:

- $Y_{(n)} = \max_{1 \le i \le n} Y_i$: the largest observed time
- $Y_{(n)}^* = \max\{Y_i : \delta_i = 1\}$: the largest uncensored observation
- $\tau_{F_0} = \sup\{t : F_0(t) < 1\}$: the right endpoint of the uncured failure-time distribution
- $\tau_G = \sup\{t : G(t) < 1\}$: the right endpoint of the censoring distribution

**Testing for a non-zero cure fraction.** For the first step, Maller and Zhou[27] proposed testing $H_0$: $p = 1$ versus $H_1$: $p < 1$, using the estimator $\hat{p}_n = \hat{F}_n(Y_{(n)})$, the Kaplan-Meier estimate evaluated at the largest observed time. They also describe a parametric version using the deviance statistic $d_n$ from a fitted parametric model[28]. For additional testing methods, see Peng and Yu[8].

**Testing for sufficient follow-up.** For the second step, Maller and Zhou[24] established that $\hat{p}_n$ is a consistent estimator of $p$ if and only if $\tau_{F_0} \le \tau_G$, and proposed testing $H_0$: $\tau_{F_0} > \tau_G$ (insufficient follow-up) versus $H_a$: $\tau_{F_0} \le \tau_G$ (sufficient follow-up). Rejecting $H_0$ indicates sufficient follow-up. The intuition is that if $\tau_{F_0} > \tau_G$ (insufficient follow-up), then $Y_{(n)} - Y_{(n)}^* \to 0$ almost surely, so a large value of $Y_{(n)} - Y_{(n)}^*$ provides evidence that $\tau_{F_0} \le \tau_G$.

Define $N_n$ as the number of uncensored observations in $(2Y_{(n)}^* - Y_{(n)}, Y_{(n)}^*]$. They formulated one test statistic as:

$$\hat{\alpha}_n = (1 - N_n/n)^n \quad (4)$$

with $H_0$ rejected (indicating sufficient follow-up) if $\hat{\alpha}_n < 0.05$. Maller and Zhou[17] noted that the $\hat{\alpha}_n$ test can have inflated type I error. To address the problem of type I error inflation, two alternative test statistics were developed by Maller and Zhou[17] and Shen[29], known as $q_n$ and $\tilde{\alpha}_n$ respectively. The $q_n$ statistic was initially limited because it required simulation-based percentiles, but this was recently resolved by Maller et al.[30]. Other authors have also recently developed alternative methodologies within this framework. Xie et al.[31] inverted the null and alternative hypotheses for sufficient follow-up. Following the logic of a different framework proposed by Selukar and Othus[32] (described next), Yuen and Musta[33] modified the null hypothesis to allow a user-specified proportion of uncured subjects to remain at risk at the analysis time.

**RECeUS Method.** Instead of the two-step framework, Selukar and Othus[32] combined testing for a non-zero cure fraction and sufficient follow-up into one procedure. Using the mixture cure framework defined in Section 2.1, let $\tau$ denote the maximum observed follow-up time.

Define $\hat{r}_n$ as the estimated uncured survival at τ divided by the estimated overall survival at τ:

$$\hat{r}_n = \hat{S}_u(\tau) / [(1 - \hat{p}_n) + \hat{p}_n \cdot \hat{S}_u(\tau)], \quad (5)$$

where the hat notation represents estimates of the corresponding population quantities. This ratio is a standardized measure of the proportion of uncured subjects still at risk at the end of follow-up. It incorporates the cure fraction and the censoring pattern.

Small values of $\hat{r}_n$ suggest sufficient follow-up for cure modeling. The proposed procedure is:

(i) Fit parametric cure models and non-cure models and compare using AIC. If a non-cure model is selected, a cure model is not appropriate.

(ii) If a cure model is selected, estimate $S(\tau)$, $S_u(\tau)$, and $(1 - p)$.

(iii) If $(1 - \hat{p}_n)$ is small, then either a cure model is likely not valid or follow-up is likely insufficient.

(iv) If $(1 - \hat{p}_n)$ is away from 0, estimate $r$. Small values of $\hat{r}_n$ indicate sufficient follow-up.

Selukar and Othus[32] recommend thresholds of $(1 - \hat{p}_n) > 0.025$ and $\hat{r}_n < 0.05$, and RECeUS claims a cure model is appropriate when both conditions hold. If one or both conditions fail, RECeUS concludes a cure model is not appropriate.

## 3 | A WORKED EXAMPLE ON ASSESSING CURE MODEL APPROPRIATENESS

We analyzed data from a phase 3 randomized trial S1203 (ClinicalTrials.gov: NCT01802333) for adult patients with previously untreated acute myeloid leukemia (AML)[34]. For presentation purposes, we summarize results for one study arm: induction with idarubicin and high-dose cytarabine followed by consolidation with cytarabine and idarubicin ("IA"). These data have maximum follow-up of approximately 7.3 years since randomization.

We analyze event-free survival (EFS), defined as time to all-cause mortality, relapse after achieving complete remission, or completion of protocol induction/re-induction therapy without achieving complete remission. EFS is measured from each patient's time of randomization, and patients without the respective events are censored at their last contact date. We use survival data from the IA arm throughout this worked example in this section to illustrate the cure-model appropriateness assessment.

The worked example is presented step by step below using code from the companion R package cureAssess, available in the CRAN repository (https://CRAN.R-project.org/package=cureAssess).

**Step 1: Clinical Rationale for a Cured Subpopulation**

In the context of AML, approximately 22.6% of adult AML patients survive to five years. While late relapses are possible, 5-year survivors are typically considered cured[35]. This provides evidence that a non-zero cure fraction in the population may be considered for this study.

**Step 2: Visual Evidence**

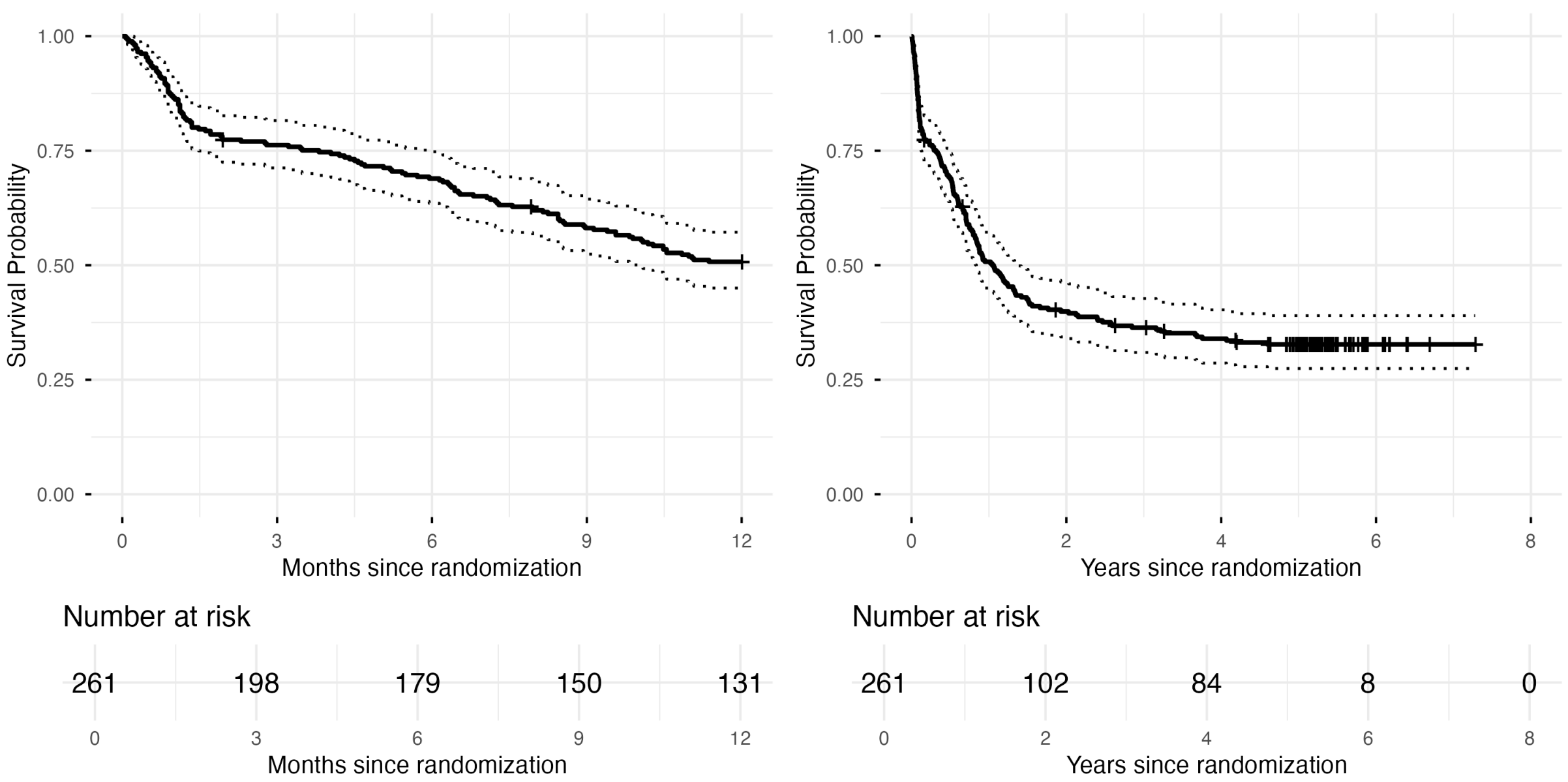

[FIGURE 2: (left) KM estimates of EFS for the IA arm after artificially restricting follow-up time to 1 year; (right) KM estimates of EFS for the IA arm for the total follow-up time]

The right panel of Figure 2 shows the Kaplan-Meier curve for the IA arm using all available follow-up (more than seven years). We observe no events for over a year before the end of follow-up, producing a long horizontal plateau, and the rate of events appears lower after year four compared to earlier. To illustrate what inadequate follow-up could look like by contrast, the left panel shows the same data after artificially restricting follow-up to one year. In this restricted view, events occur shortly before one year, the curve continues to step downward up to the end of the period, and no horizontal plateau is visible.

From this visual comparison, the right panel suggests that patients remaining at risk after seven years are unlikely to experience an EFS event with further follow-up, supporting the plausibility of a cure model. Conversely, an observer would likely be skeptical of a similar claim from the left panel alone. We therefore conclude that a cure model may be appropriate when considering the full available follow-up but would likely not be appropriate if the study follow-up were restricted to one year.

**Step 3: Quantitative Evidence**

For illustrative purposes, we conducted two approaches: (1) the two-step approach using the Maller–Zhou test for non-zero cure fraction and the $q_n$ test[30] for sufficient follow-up; and (2) the RECeUS method[32]. The two-step approach suggests a non-zero cure fraction and sufficient follow-up and the RECeUS method also suggests strong evidence for cure modeling of the IA arm for EFS. We provide worked calculations for each of these approaches below.

*Two-step approach*

*Test for a non-zero cure fraction.* Following the first step of the two-step procedure, we test $H_0$: $p = 1$ (no cure fraction) against $H_1$: $p < 1$ using $\hat{p}_n = \hat{F}_n(Y_{(n)})$, the Kaplan–Meier estimate of the event probability at the largest observed time. For the IA arm the largest observed EFS time (7.28 years) is censored, and the Kaplan–Meier curve levels off at $\hat{S}(Y_{(n)}) = 0.327$, so $\hat{p}_n = 1 - 0.327 = 0.673$, with 87 of 261 subjects (33.3%) censored. A formal test of $H_0$: $p = 1$ compares $\hat{p}_n$ against critical values (indexed by sample size and censoring proportion). Here $\hat{p}_n = 0.673$ lies far below 1 and well below these 5% critical values, so we reject $H_0$ and conclude that a non-zero cure fraction is present.

```
library(cureAssess)
ia  <- prepare.surv.data(ia_arm, time = "efs_time", status = "efs_event",
                          time_scale = "days_to_years")
```

```
immune.test(ia)
```

*Test for sufficient follow-up.* For the second step we use the $q_n$ test following Maller et al.[30] The largest observed time is $Y_{(n)}$ = 2659 days = 7.28 years (censored) and the largest event time is $Y_{(n)}$* = 1679 days = 4.60 years, so the late-time window ($2Y_{(n)}* - Y_{(n)}$, $Y_{(n)}$*] = (699, 1679] days contains $N_n$ = 19 events. The test statistic is $q_n = N_n/n$ = 19/261 = 0.073. We reject $H_0$ (insufficient follow-up) at the 5% level when $n \cdot q_n = N_n$ exceeds the asymptotic critical value log(0.05)/log(3/4) − 1 ≈ 9.41; here $n \cdot q_n$ = 19 > 9.41, so we conclude that follow-up is sufficient. Taken together, the two-step approach finds evidence of both a non-zero cure fraction and sufficient follow-up.

```
qn.test(ia)
```

*RECeUS approach*

We first fit a class of parametric cure and non-cure models and use AIC to select an appropriate model. (Importantly, this model for assessing cure model appropriateness can differ from the analysis model.) These models are fit with the cureAssess package. We first put the IA-arm data into the format the package expects using prepare.surv.data(): this creates a survival-time column Y and an event indicator D (1 = event, 0 = censored) and, with time_scale = "days_to_years", rescales follow-up from days to years by dividing by 365.25. We then call model.fitting(), which fits a mixture-cure and a non-cure model for each of the exponential, Weibull, gamma, and log-logistic distributions and orders them by AIC.

```
library(cureAssess)
ia  <- prepare.surv.data(ia_arm, time = "efs_time", status = "efs_event",
                         time_scale = "days_to_years")
fit <- model.fitting(ia)
fit$aic_table          # AIC comparison shown in Table 1
```

Table 1 reports the exponential, Weibull, gamma, and log-logistic families, which make up the model class employed for the RECeUS method in Selukar et al.; the cureAssess model.fitting() function additionally allows lognormal cure and non-cure models.

The Weibull cure model has the smallest AIC value for the IA arm, so we proceed to estimate $(1 - \hat{p}_n)$ and $\hat{r}_n$. The Weibull mixture cure model gave an estimated cure fraction $(1 - \hat{p}_n) = 0.3217$ (95% CI, 0.2663–0.3826), with Weibull shape = 0.8046 and scale = 0.7535, from 261 patients (174 events and 87 censored). The maximum follow-up is $\tau = 2659$ days / 365.25 = 7.28 years. Based on these elements, we can calculate the necessary quantities for RECeUS.

Calculate the survival function for susceptibles at $\tau$:

$$\hat{S}_u(\tau) = \exp[-(\tau/scale)\text{^}shape] = \exp[-(7.28/0.7535)\text{^}0.8046] = 0.002024$$

Calculate the population survival at $\tau$:

$$\hat{S}(\tau) = (1 - \hat{p}_n) + \hat{p}_n \times \hat{S}_u(\tau) = 0.3217 + (0.6783)(0.002024) = 0.3231$$

Calculate the ratio:

$$\hat{r}_n = \hat{S}_u(\tau) / \hat{S}(\tau) = 0.002024 / 0.3231 = 0.00626$$

These quantities are also returned directly by cureAssess. Applying receus.method(ia, dist = "wei") to the Weibull cure model returns the estimated cure fraction $(1 - \hat{p}_n)$, the RECeUS ratio $\hat{r}_n$, the two threshold checks, and an overall decision. By default, the evaluation time $\tau$ is the maximum observed follow-up. For the IA arm it returns $(1 - \hat{p}_n) = 0.3217$ and $\hat{r}_n = 0.00626$, matching the values above.

Since $(1 - \hat{p}_n) = 0.3217 > 0.025$ and $\hat{r}_n = 0.00626 < 0.05$, both conditions are satisfied. Therefore, RECeUS confirms that a cure model is appropriate for this dataset.

The entire IA-arm assessment above can also be reproduced in a single call with the cureAssess wrapper cure.appropriateness(), which formats the data, compares the candidate cure and non-cure models by AIC, and runs the diagnostic tests to return an overall recommendation:

```
cure.appropriateness(
  data =ia_arm,
  time = "efs_time",
  status = "efs_event",
  time_scale = "days_to_years",
  plot_km = FALSE,
  run_tests = "yes"
)
```

## 4 | OTHER PRACTICAL CONSIDERATIONS FOR APPROPRIATE APPLICATION OF CURE MODELS: BMTCT DATA EXAMPLES

To share broader considerations for appropriately applying cure models, we applied the assessment framework to a series of datasets from Bone Marrow Transplantation & Cellular Therapy (BMTCT) research at St. Jude Children's Research Hospital. These included both prospective clinical trials and retrospective analyses of hematopoietic cell transplantation (HCT). HCT is considered a potentially curative treatment option for certain patient populations, including pediatric patients with high-risk hematological malignancies.

The goals of this work were to evaluate potential cure model appropriateness across (1) different patient populations and (2) varying follow-up procedures. For goal (1) we included BMTCT cohorts with different expected prognoses (e.g., patients in complete remission [CR] at the time of transplantation are expected to have better prognosis than patients with active disease at the time of transplantation). For goal (2) we included both prospective trials (with active, protocol-specified monitoring) and retrospective analyses (where follow-up may depend on routine clinical documentation rather than protocol-specified visits), allowing the appropriateness framework to be evaluated across the kinds of follow-up regimes investigators encounter in practice.

The prospective datasets comprised two studies of high-risk hematological malignancies for patients with no prior transplant and containing multiple prespecified cohorts: HAPNK1 [36] (phase 2 study [NCT01807611] with a CR cohort, $n = 53$ and active disease cohort, $n = 19$) and HAP2HCT (phase 1/2 study [NCT03849651]; here, we analyzed phase 2 patients at dose level 3, $n = 48$).

The retrospective datasets comprised two analyses, also with multiple prespecified cohorts: Epperly et al.[37] (“HCTRETRO”, patients receiving ≥2 HCTs; with a 2nd HCT cohort, $n = 106$; and a >2nd HCT cohort, $n = 13$) and Talleur et al. (“Refractory at HCT”, [unpublished] patients with refractory disease at transplant, $n = 129$).

The complete table of cure model appropriateness assessments for all datasets is provided in the Supplementary Materials (Table S1). Here we present two illustrative examples.

*Example: HCTRETRO 2nd HCT Cohort*

This cohort ($n$ = 106) had median follow-up of 0.84 years and maximum follow-up of 17.7 years. The Kaplan-Meier estimate at maximum follow-up was 0.257. Visual inspection showed a plateau in the KM curve consistent with long-term survivors, and AIC comparison selected the Weibull cure model. Then, RECeUS assessment confirmed cure model appropriateness. This example illustrates how retrospective data with extended follow-up can support cure model application despite the lack of follow-up in the context of a clinical research protocol.

*Example: HAP2HCT*

The phase 2 portion of this study included 48 patients with a maximum follow-up of 1.6 years. Using the maximum follow-up, there was limited visual evidence for a cure fraction and a Kaplan-Meier estimate at maximum follow-up of 0.277. The clinical trial had 1 year of active follow-up, so we hypothesized that follow-up after 1 year could be sensitive to differing reporting quality when switching from active to passive follow-up. In a sensitivity analysis administratively censoring at 1 year, conclusions for cure model appropriateness reversed with both visual and quantitative evidence supporting cure modeling. This demonstrates a potential impact of active versus passive follow-up when employing cure models.

## 5 | DISCUSSION

It is important for cure models to be clinically meaningful and appropriately tailored to the specific data being analyzed. While potentially useful tools, applying cure models when conditions are not met can lead to misleading results. In this tutorial, we provide guidance to assess the appropriateness of cure models through a combination of expert judgment, visual assessment, and quantitative evaluation. First, expert knowledge determines whether a cure is biologically or clinically plausible. Next, visual inspection of the Kaplan-Meier curve can reveal a plateau that does not approach a survival probability of zero, suggesting the presence of long-term survivors. Finally, statistical methods can provide quantitative evidence to confirm the appropriateness of cure models.

Recent work by Selukar and Othus[32] and Yuen and Musta[33] proposed that cure models may still provide useful inference even when the classical sufficient follow-up condition $\tau_{F_0} \leq \tau_G$ is not strictly satisfied. Specifically, these methods suggest cure models may be appropriate when a small but non-negligible proportion of susceptible individuals remain at risk at the end of follow-up. While their simulation studies support this approach, the theoretical justification remains an area of ongoing research. There is no universally accepted threshold for how much the condition can be relaxed while still obtaining valid estimation and inference.

This tutorial evaluates one group at a time, as this has been the focus for each method proposed thus far. Kouadio et al.[38] investigated cure model appropriateness for randomized controlled trials and suggest that appropriateness in at least one of two arms could be sufficient for applying cure models. Importantly, in their general proofs of the large-sample properties of cure models, Patilea and Van Keilegom[22] assume that cure model appropriateness holds at every point of the covariate space, so extending this to models with more than two groups may be more difficult. Evaluating appropriateness for each possible group may be challenging due to multiplicity and sample size limitations, so special considerations may be required.

Through our BMTCT data examples, we identified other practical considerations for applying cure models. Cure model appropriateness cannot be predicted solely by the quantity of follow-up: we identified scenarios in which highest-risk populations with suitable follow-up showed evidence that they could potentially be appropriately modeled with cure models, whereas lower-risk populations may not, even with a longer duration of follow-up. We also observed potential data issues for cure modeling when follow-up transitioned from active trial monitoring to passive follow-up, especially with very long maximum follow-up.

## 6 | CONCLUSION

In summary, while cure models offer valuable insights for clinical trials, their implementation requires careful attention to data and study considerations such as follow-up time, data quality, and model appropriateness. By ensuring clinical relevance and assessing these factors with the systematic procedure described in this tutorial, researchers can enhance the utility and reliability of cure models, supporting more effective analyses and improved patient outcomes.

## ACKNOWLEDGEMENTS

This work was supported in part by funding from the American Lebanese Syrian Associated Charities and National Cancer Institute grant awards U10CA180888 and U10CA180819. This work used data from SWOG clinical trial S1203 (ClinicalTrials.gov: NCT01802333), with primary trial results reported by Garcia-Manero et al. The bone marrow transplantation and cellular therapy datasets were provided by St. Jude Children's Research Hospital. We thank both organizations for providing access to the data used in this work.

## AUTHOR CONTRIBUTIONS

Subodh Selukar conceived the project, provided supervision and guidance throughout all steps, and contributed scientific feedback. Geethanjalee Mudunkotuwa and Durbadal Ghosh developed the code and the cureAssess R package and performed the analyses. Brandon Triplett provided clinical input and scientific feedback. Geethanjalee Mudunkotuwa, Durbadal Ghosh, and Subodh Selukar drafted the manuscript, and all authors reviewed the manuscript. Geethanjalee Mudunkotuwa and Durbadal Ghosh contributed equally to this work.

## CONFLICT OF INTEREST STATEMENT

The authors declare no conflicts of interest.

## DATA AVAILABILITY STATEMENT

The SWOG S1203 trial data (ClinicalTrials.gov NCT01802333) analyzed in Section 3 may be requested through the SWOG request process. The BMTCT datasets analyzed in Section 4 may be requested from the corresponding author, subject to institutional approval. All analyses used the authors' R package cureAssess, which is available on CRAN (https://CRAN.R-project.org/package=cureAssess).

## USE OF ARTIFICIAL INTELLIGENCE TOOLS

Artificial intelligence (AI) tools were used only to polish the English language of the manuscript and to help debug and verify computer code written by the authors; they were not used for any other purpose. The authors reviewed all AI-assisted output and take full responsibility for the content of this article.

**Table 1: Comparing parametric mixture cure models and non-cure models**

| Model | IA Arm AIC |
|---|---|
| Exponential non-cure | 768.52 |
| Exponential cure | 599.45 |
| Weibull non-cure | 633.33 |
| Weibull cure | 588.80* |
| Gamma non-cure | 651.01 |
| Gamma cure | 590.56 |
| Log-logistic non-cure | 610.08 |
| Log-logistic cure | 590.45 |

*Lowest AIC